\documentclass[10pt]{amsart}
\usepackage{amsmath,amssymb,amsthm,amscd,mathtools, graphicx, mathtext,color,wrapfig}
\usepackage[numbers,sort&compress]{natbib}
\usepackage{epsfig}
\usepackage{courier}
\begin{document}


\section*{String Theory and Math: Why This Marriage May Last}

{\centerline{\it Mathematics and Dualities of Quantum Physics}}

\vspace{1pc}
{\centerline{Mina Aganagic}}
\vskip 0.13cm 
\vskip 0.13cm 

\begin{abstract}
String theory is changing the relationship between mathematics and physics. I will try to explain how and why. The central role is played by the phenomenon of duality, which is intrinsic to quantum physics and abundant in string theory.
\end{abstract}

The relationship between mathematics and physics has a long history. Traditionally, mathematics provides the language physicists use to describe Nature, while physics brings mathematics to life: To discover the laws of mechanics, Newton needed to develop calculus. The very idea of a precise physical law, and mathematics to describe it were born together. To unify gravity and special relativity, Einstein needed the language of Riemannian geometry. He used known mathematics to discover new physics. General relativity has been inspiring developments in differential geometry ever since. Quantum physics impacted many branches of mathematics, from geometry and topology to representation theory and analysis, extending the pattern of beautiful and deep interactions between physics and mathematics throughout centuries.

\vskip 0.13 cm
String theory brings something new to the table: the phenomenon of duality. Duality is the equivalence between two descriptions of the same quantum physics in different classical terms.  Ordinarily, we start with a classical system and quantize it, treating quantum fluctuations as small.
However, nature is intrinsically quantum. One can obtain the same quantum system from two distinct classical starting points. For every precise question in one description of the theory, there is a corresponding question in the dual description.
Duality is similar to a change of charts on a manifold, except it also has the power to
map large fluctuations in one description to small fluctuations in the dual, and relate very hard mathematical problems in one are of mathematics to more manageable ones in another. Dualities are pervasive in string theory.
\vskip 0.13 cm
Understanding dualities requires extracting their mathematical predictions and proving the huge set of mathematical conjectures that follow. The best understood duality is {\it mirror symmetry}. But, mirror symmetry is but one example -- many striking dualities have been discovered in quantum field theory (QFT) and many more in string theory over the last 20 years. Duality gives quantum physics, and especially string theory, the power to unify disparate areas of mathematics in surprising ways and provides a basis for a long lasting and profound relationship between the physics and mathematics.\footnote{Reviews of dualities in string theory can be found in \cite{Polchinski:1996nb,
Vgp, polchinski}. For another review of interaction between mathematics and physics see \cite{moore}.}


\section{Knot theory and Physics}

To illustrate these ideas, I will pick one particular area of mathematics, knot theory. The central question of knot theory is: When are two knots (or links) distinct? A knot is an oriented closed loop in  ${\mathbb R}^3$. A link consists of several disjoint, possibly tangled knots. Two knots are considered equivalent, if they are homotopic to each other. One approaches the question by constructing knot or link invariants, which depend on the knot up to homotopy. 

Knot theory was born out of 19th century physics. Gauss' study of electromagnetism resulted in the first link invariant: the Gauss linking number, which is an invariant of a link with two knot components, $K_1$ and $K_2$. One picks a projection of the link onto a plane and defines (twice) the linking number as the number of crossings, 
\begin{figure}[h]
\includegraphics[trim = 1mm 50mm 1mm 50mm, clip,
 width=6cm]{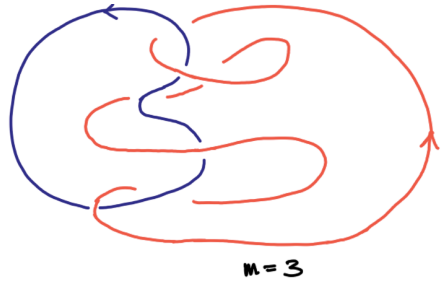}
\caption{
To define the sign of a crossing, we approach a crossing along the bottom strand, and assign $+1$ if the top strand passes from left to right, and $-1$ otherwise. In the figure, there are 6 crossings, each contributing $+1$, and so $m=3$.}
\label{one}
\end{figure}
counted with signs:

\begin{align}\label{first}
m(K_1, K_2)= {1\over 2} \sum_{{\rm crossings}(K_1, K_2)} {\rm sign}({\rm crossing}).
\end{align}
Gauss discovered the linking number, and gave a beautiful integral formula for it:
\begin{align}\label{second}
m(K_1, K_2)= {1\over 2 \pi}\oint_{K_1} \oint_{K_2}{{\vec x}_1-{\vec x}_2\over |{\vec x}_1 -{\vec x}_2|^3} \cdot ( d{\vec x}_1 \times d {\vec x}_2).
\end{align}    
Maxwell discovered it independently, some time later, and noted that it is not a very good link invariant -- it is easy to find links that are non-trivial, yet the invariant vanishes.  Note that, while the first formula \eqref{first} for the linking number relies on a choice of a projection, the second one \eqref{second} makes it manifest one is studying a link in three dimensional space.
 
Strikingly, quantum physics enters knot theory. In $'84$, Vaughan Jones found a very good polynomial invariant of knots and links, by far the best at the time, depending on one variable $q$ \cite{jones1, jones2}. The Jones polynomial is a Laurent polynomial in $q^{{1\over 2}}$; it can be computed in a simple way by describing how it changes as we reconnect the strands and change the knot. One picks a planar projection of the knot, and a neighborhood of a crossing. 
\begin{figure}[h]
\includegraphics[trim = 1mm 90mm 1mm 70mm, clip, width=9cm]{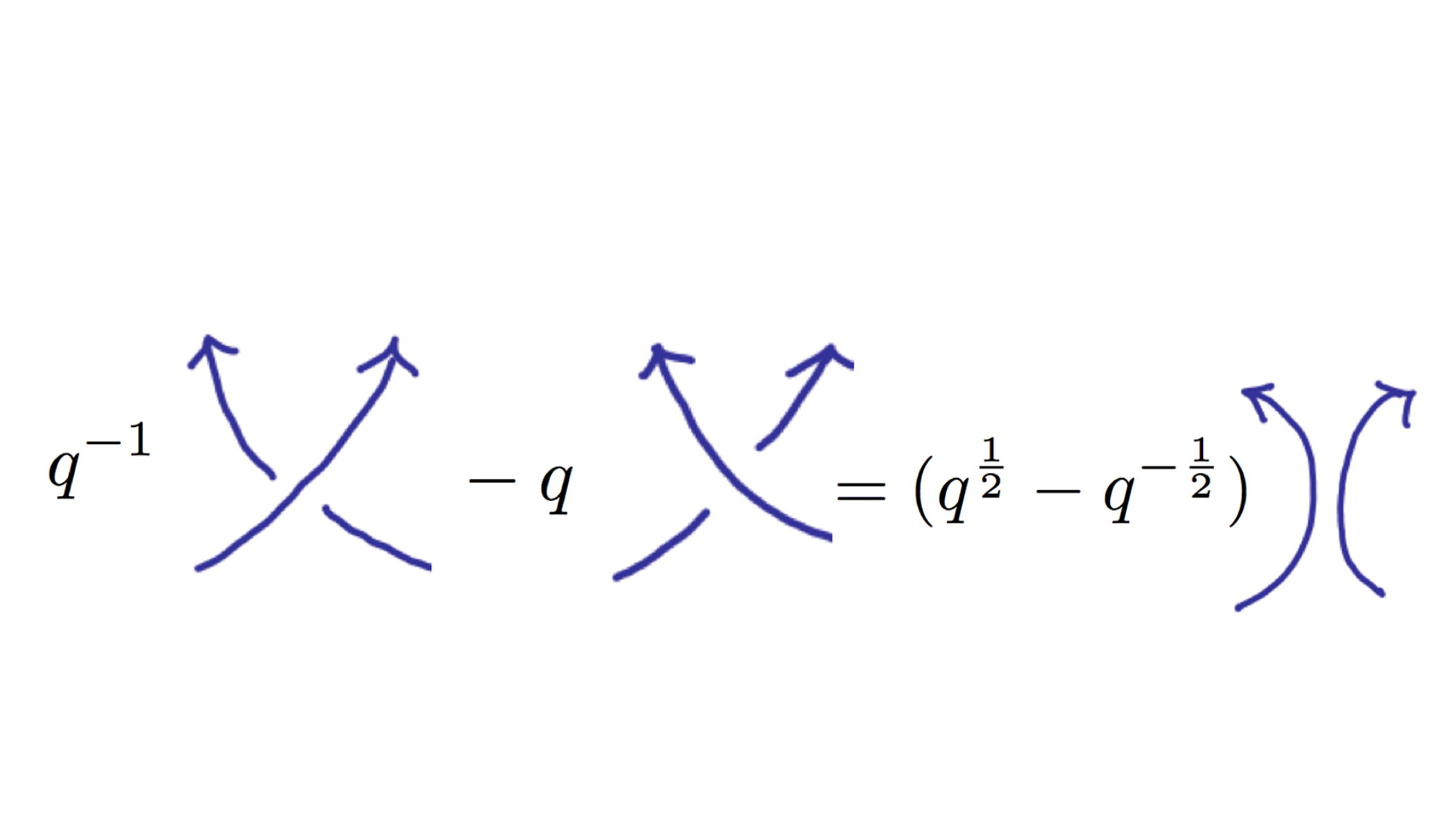}
\caption{Skein relation for the Jones polynomial}
\label{two}
\end{figure}
and defines the value of the Jones polynomial by the skein relation it satisfies:

$$q^{-1} J_{K_+} - q J_{K_-} = (q^{1\over 2} - q^{-{1\over 2}} )J_{K_0}.
$$
together with specifying its value for the unknot. 
While there are examples of distinct knots with the same $J_K(q)$, there is no known examples of non-trivial knots with $J_K(q)$ the same as for the unknot. Despite the ease of construction, the Jones polynomial seems mysterious. Since one has to pick a projection to a plane to define it, it is not obvious at the outset that one obtains an invariant of knots in three dimensional space, rather this is something one must prove. Secondly, what is the meaning of $q$?

Witten discovered that the Jones polynomial has its origin in quantum field theory: Chern-Simons (CS) gauge theory in three dimensions. Like Yang-Mills theory, Chern-Simons theory on a three-manifold $M$ is written in terms of a connection
$$ A = A_i dx^i
$$
associated with a gauge group $G$. The theory is topological from the outset -- its classical action is given in terms of Chern-Simons form on $M$,
$$
S_{CS} = {1\over 4 \pi} \int_M {\rm Tr}(A \wedge dA + {2\over 3} A \wedge A \wedge A).
$$
and hence it is independent of the choice of metric on $M$.
The path integral of the theory
$$
Z(M) = \int {\mathcal  D} A \, {\rm exp}\Bigl(i k S_{CS}\Bigr),
$$
where one integrates over spaces of all connections on $M$ and divides by the gauge group is a topological invariant of $M$. 
We can introduce a knot $K$ in the theory by inserting a line observable along $K$,
$$
{\mathcal  O}_K(R) = {\rm Tr}_R \,Pexp \Bigl( i\oint_K A_i dx^i\Bigr)
$$
in some representation $R$ of the gauge group ($P$ denotes path ordering of the exponential). This preserves topological invariance, so
$$
Z(M;K, R)= \int {\mathcal D} A \; {\rm exp}(i k S_{CS}) \;{\it O}_K(R)
$$
is a topological invariant of the knot $K$ in the three manifold $M$, which depends only on $G$, $R$ and $k.$
(More precisely, Chern-Simons theory produces an invariant of a framed three-manifold $M$, with framed knots. Framing is a choice of a homotopy class of trivialization of the tangent bundle of $M$ and $K$.  The need to fix the framing reflects an ambiguity in the phase the partition function \cite{WJ}.) The constant $k$ is required to be an integer, for the integrand to be invariant under "large" gauge transformations, those corresponding to non-trivial elements of ${\mathcal \pi}_3(G)$.

Witten made use of the topological invariance of the theory to solve Chern-Simons theory exactly on an arbitrary three manifold $M$ with collection of knots, by cutting the three manifold into pieces, solving the theory on pieces and gluing back together. He showed that, taking 
$$M=S^3, G = SU(2), R=\Box,
$$
where $R=\Box$ is the defining two dimensional representation of $SU(2)$, a suitably normalized Chern-Simons partition function
$$
\langle {\mathcal  O}_K \rangle=Z(M;K)/Z(M; \bigcirc)
$$
equals the Jones polynomial
$$
 \langle {\mathcal  O}_K  \rangle = J_K(q).
 $$
The normalization we chose corresponds to setting $J_{\bigcirc}(q)=1$. Chern-Simons theory gave a manifestly three dimensional formulation of the Jones polynomial. It leads immediately to a vast generalization of Jones' knot invariant, by varying the gauge group $G$, the representations, and considering knots in an arbitrary three manifold $M$. 
Finally, the relation to Chern-Simons theory showed that the Jones polynomial is a quantum invariant: it is a Laurent polynomial in 
$$q= exp(i\lambda),$$ 
where $\lambda = 2\pi /(k+2)$ plays the role of  $\hbar$, the Planck constant, in Chern-Simons theory.

Let me pause for a moment to sketch what one means by saying the theory is solvable \cite{WJ}. It is known that every three manifold $M$ can be related to $S^2\times S^1$ by a repeated application of surgery. A surgery to produce from $M$ a new three manifold proceeds as follows.
One picks an imaginary knot in $M$, cuts out its solid torus neighborhood, and glues it back in up to an $U\in SL(2,Z)$ transformation of the boundary. If $U$ is not identity one obtains a new manifold $M'$. Quantum field theory is a functor that associates to a closed three manifold $M$ a complex number $Z(M)$, the value of the path integral on $M$, and to a manifold with a boundary $B$ a state in the vector space ${\mathcal H}_B$,  the Hilbert space of the theory based on $B$. Vector spaces associated to the same $B$, with opposite orientation, are canonically dual. Gluing two manifolds over a common boundary $B$ is the inner product of the corresponding states. So surgery on three manifolds translates to a following statement in QFT:
$$
Z(M') = \langle 0  | M' /K\rangle = \langle 0  | U|M/K\rangle
$$
Here $\langle 0| $ is the state corresponding to solid torus with no insertions, and $| M' /K\rangle$ the state corresponding to $M'$ with a neighborhood of the knot $K$ cut out. An arbitrary state in ${\mathcal H}_{T^2}$ can be obtained from a solid torus with a line observable colored by a representation $R$ running through it.  If we denote the resulting state $\langle R|$, we can write
$$
 \langle 0  | U = \sum_R \langle R| \, U_{0R}.
$$
The sum runs over a finite set of representations of $G$, depending on $k$. (The Hilbert space ${\mathcal H}_{B}$ of Chern-Simons theory with gauge group $G$ at level $k$ is the same as the space of conformal blocks of $G_k$ WZW model; the latter is finite dimensional for any $B$.)
This implies
 $$
Z(M') = \sum_R \langle R| M/K\rangle \;  U_{0R}= \sum_R Z(M', K, R)\;  U_{0R}
 $$
where $Z(M', K, R)$ corresponds to the partition function on $M'$ with an actual knot $K$ colored by representation $R$ running through it. In this way, by repeated surgeries, we can reduce any three manifold invariant to that of $S^2\times S^1$ with a braid running along the $S^1$. In turn, the later can be computed by
$$
Z(S^2\times S^1, L, R_i) = {\rm  Tr}_{\,{\mathcal H}_{S^2, R_i}}B_L,
$$
which comes about by first cutting the $S^2\times S^1$ open into $S^2 \times R$, straightening the braid out, and then recovering the original braid by finding a collection $B_L$ of time-ordered diffeomeorphisms of a sphere $S^2$ with marked points, which re-braid the braid. Gluing the ends together corresponds to taking the trace of $B_L$, acting on the Hilbert space ${\mathcal H}_{S^2, R_i}$ of the theory on $S^2$ with marked points colored by representations $R_i$ determined by the braid.

To solve the theory one needs only a finite set of data. The $SL(2,\mathbb{Z})$ transformations of the torus are generated by a pair of matrices, $S$ and $T$ satisfying
$$S^4=1,\qquad  (ST)^3=S^2,
$$ 
representing the action of $SL(2, {\mathbb Z})$ on ${\mathcal H}_{T^2}$. Similarly, the brading matrix $B_L$ from is obtained from a finite set of data, the braiding matrix $B$ and fusion matrix $F$ on a four punctured sphere \cite{RT}. For Chern-Simons theory based on gauge group $G$, at level $k$, the $S$, $T$, $B$ and $F$ are finite dimensional matrices acting on conformal blocks of$G_k$ WZW 2d CFT.  
Reshetikhin and Turaev formalized this in terms of modular tensor categories \cite{RT}. 
Thus, one can reduce finding knot and three manifold invariants for arbitrary $G$ and $k$ and representations $R$ to matrix multiplication, of a small set of matrices.

\section{Gromov-Witten Theory}

Quantum physics enters modern mathematics in other places as well. 
\emph{Gromov-Witten theory} is an example. There, one studies quantum intersection theory of a projective variety $X$ (see \cite{M} for a review, and \cite{Okounkov} for a quick overview).
Classical intersection corresponds to picking classes
$$
\gamma_1, \ldots ,\gamma_n \in H_*(X)
$$
with degrees $\sum_i {\rm deg} (\gamma_i^\vee) = 2d$, where $d={\rm dim}_{\mathbb C}(X)$ and computing their intersection numbers, counted with signs:
\begin{equation}\label{classc}
\langle \gamma_1, \ldots , \gamma_n\rangle_{0,0}= \int_X\gamma_1^{\vee}\wedge\cdots \wedge \gamma_n^\vee,
\end{equation}
where $\gamma_i^\vee\in H^*(X)$ denotes the Poincare dual of $\gamma_i$.
Enumerative geometry turns this into a deeper geometric question by counting intersections over algebraic curves, insead over points: one would like to know how many algebraic curves of a give degree $\beta\in H_{2}(X)$ and genus $g$ meet $\gamma_1, \ldots, \gamma_n $ at points. The corresponding invariant 
$$
\langle \gamma_1, \ldots , \gamma_n\rangle_{g,\beta}.
$$
can be defined by picking a curve $\Sigma$ of genus $g$, with $n$ marked points $p_1$, \ldots $p_n$, and considering intersection theory on the moduli space ${\mathcal M}_{g, n}(X, \beta)$ of holomorphic maps 
$$
\phi: \Sigma \rightarrow X
$$
of degree $d$. More precisely, as explained by Kontsevich, one needs consider \emph{moduli space of stable maps}  ${\overline {\mathcal M}_{g, n}(X)}$.  This is a compactification of  ${\mathcal M}_{g, n}(X, \beta)$ by  allowing the domain curve ${\Sigma}$ to have "ears", which are additional $S^2$ that bubble off, and considering \emph{ stable maps}, which he defined. Imposing the incidence condition that $\phi(p_i) \subset \gamma_i$ is implemented by pulling back the Poincare dual class $\gamma_i^\vee$ via the evaluation map ${\rm ev}_i$. The evaluation map maps a point in the moduli space of maps to $\phi(p_i)$:

\begin{equation}\label{qint}
\langle \gamma_1, \ldots , \gamma_n\rangle_{g,d} = \int_{[\overline{{\mathcal M}}_{g, n}(X, \beta)]} \;{\rm ev}_1^*(\gamma_1^{\vee})\cdots {\rm ev}_n^*(\gamma_n^{\vee}),
\end{equation}
where the brackets $[..]$ denote the (virtual) fundamental class.  For genus zero, degree zero curves, the definition agrees with the classical intersection numbers in \eqref{classc}. At genus zero, it is natural to combine the classical answer \eqref{classc} and the higher degree data into a generating function of quantum intersection numbers of $X$,

\begin{equation}\label{qint2}
\langle \gamma_1, \ldots , \gamma_n\rangle_{0,Q} =  \sum_{\beta \in H_{2}(X) } \langle \gamma_1, \ldots , \gamma_n\rangle_{0,\beta}\, Q^\beta.
\end{equation}
For a map of degree $\beta$ to $X$, $Q^\beta$ is the exponent the area of the target curve, $Q^\beta = exp(- \int_{\Sigma} \phi^*\omega)$, where $\omega$ is the Kahler form on $X$.
The leading term in the series is the classical intersection, and the subleading terms are quantum corrections to it.

\subsection{Gromov-Witten Theory and Topological String Theory}
Gromov-\linebreak Witten theory originates from string theory. It computes the amplitudes of a topological variant of superstring theory, called the \emph{A-model topological string}.  

In quantum field theory, to describe a particle propagating on a manifold $X$ one sums over all maps from graphs $\Gamma$ to $X$, satisfying certain conditions, where one allows moduli of graph to vary.  In string theory, we replace point particles by strings, the maps from graphs $\Gamma$ by maps from Riemann surfaces $\Sigma$ to $X$. In superstring theory, one formulates this in terms of a path integral of a supersymmetric 2d QFT on $\Sigma$, describing a string propagating on $X$. To get topological string theory one modifies the supersymmetry generator $Q$ to square to zero, $Q^2=0$ on arbitrary $\Sigma$. This turns the 2d QFT into a topological quantum field theory on $\Sigma$ of cohomological type, with differential $Q$. The world sheet path integral receives contributions only from configurations that are annihilated by $Q$. If $X$ is a Calabi-Yau manifold, there are are two inequivalent ways to obtain a TQFT, leading to topological $A$- and the $B$-model string theories. They correspond to two distinct generators $Q_A$ and $Q_B$, each satisfying $Q_{A,B}^2=0$. Topological $A$-type string exists for any Kahler manifold $X$. Restricting to configurations annihilated by $Q_A$ turns out to restrict one to studying holomorphic maps to $X$ only, leading to Gromov-Witten theory. 
In the B-model, the maps annihilated by $Q_B$ are the constant maps, resulting in a simpler theory, depending on complex structure of $X$ only.

Topological string theory was introduced by Witten in \cite{Wt1,Wt2}, and developed by many (see for e.g. \cite{Wm, BCOV} and \cite{M} for a review). 
\begin{figure}[h]
\includegraphics[trim = 1mm 70mm 1mm 70mm, clip, width=8cm]{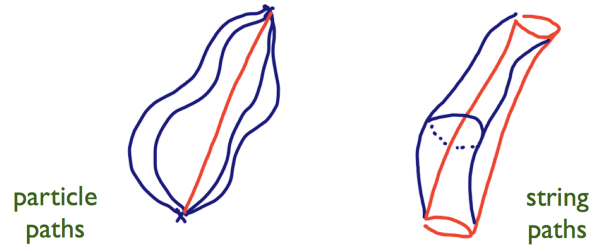}
\caption{In string theory one sums over all possible paths of a string, leading to sum over surfaces.}
\label{three}
\end{figure}
The mathematical formulation of Gromov-Witten theory is due to  Kontsevich, Manin, Fukaya and many others \cite{M}. The development of Gromov-Witten theory is an example of new mathematics that is inspired by questions in physics.

\section{Duality}

A quantum system is described by a collection of observables ${\mathcal O}_i$, corresponding to physical quantities in the theory, and expectation values of these observables,

\begin{equation}\label{correlation}
\langle {\mathcal O}_1 \ldots {\mathcal O}_n\rangle,
\end{equation}
which physicists call amplitudes, or correlation functions. In Chern-Simons theory, the observables ended up associated to knots in a three manifold $M$, colored by representations $R$ of the gauge group; in Gromov-Witten theory, the observables were related to homology classes $\gamma_i$ in $X$. 

To find the correlation functions, one starts with a classical limit of the system, and a quantization procedure. 
In Gromov-Witten theory of $X$, one would start with a two dimensional topological theory on a genus $g$ Riemann surface $\Sigma$, based on maps to $X$ \cite{M}. The description we are giving assumes quantum fluctuations are small. This opens up a possibility for the same physical system to have different descriptions, with different starting classical points, yet which result in the same set of quantum amplitudes. This expresses the fact that physics is intrinsically quantum -- only our descriptions of it rely on classical limits; and, the classical limits need not be unique. The map between the two descriptions of the single physical system, is called a \emph{duality}. 

\subsection{Mirror symmetry}
Perhaps the best known example of a duality is mirror symmetry. Mirror symmetry relates topological A-model string on a Calabi-Yau $X$, to topological B-type string theory on the mirror Calabi-Yau $Y$ (The phenomenon was discovered in \cite{VLW}, for a review see \cite{M}). The underlying Calabi-Yau manifolds are different, even topologically, as mirror symmetry reflects the hodge diamond:

$$h^{p,q}(X) = h^{d-p,q}(Y),$$
yet, the A-model on $X$ and the B-model on $Y$ are the same quantum theory. 
Here $d$ is the complex dimension of $X$ and $Y$. The amplitudes of the A-type topological string are computed by Gromov-Witten theory. The B-type topological string is reduces to a quantum field theory on $Y$ which quantizes the variations of complex structures; this is related to the fact that $Q_B$ vanishes on constant maps. In particular, the $g=0$ amplitudes can be read off from classical geometry.  The most interesting case is $d=3$, otherwise many amplitudes vanish on dimension grounds. This can be seen from the formula for the (virtual) dimension of the moduli space of stable maps in Gromov-Witten theory: ${\rm dim}_{\mathbb C} \overline{{\mathcal M}}_{g, n}(X, \beta) = -\beta \cdot c_1(TX) + (1-g)(d - 3)+n$. In the Calabi-Yau case, per definition, $c_1(TX)$ vanishes in cohomology; for $d=3$, the moduli space has positive dimension for any $g, d$ and $n$. 

The first prediction of mirror symmetry is that the genus zero amplitudes on $X$ and $Y$ agree. On $X$, computing this leads to quantum intersection numbers: choosing $\gamma_i, \gamma_j, \gamma_k$ to be three divisors in $X$, $\gamma_{i,j,k}^\vee \in H^2(X)$,  one computes
$$
\langle \gamma_i ,\gamma_j, \gamma_j\rangle_{0,Q} = \sum_{\beta \in H_2(X) } \langle \gamma_i, \gamma_j , \gamma_k\rangle_{0,\beta}\, Q^\beta.
$$
The $\beta=0$ term in the sum is the classical intersection number of the three divisor classes, and subsequent terms involve intersection theory on moduli space of stable maps, as we described above. In the mirror B-model on $Y$, the entire sum is captured \cite{COS} by \emph{classical geometry of Y}:

$$
\langle \gamma_i ,\gamma_j, \gamma_j\rangle_{0,Q} = \int_Y { \Omega} \wedge {\partial\over \partial {t_i}} {\partial\over \partial {t_j}} {\partial\over \partial {t_k}}  \; \Omega.
$$
This leads to a striking simplification. Here, $\Omega \in H^{(3,0)}(Y)$ is the unique holomorphic volume form on $Y$, whose existence is guaranteed by the Calabi-Yau condition. The parameters $t_i$ are suitably chosen moduli of complex structures on $Y$. 

The higher genus amplitudes in the $B$-model  quantize the variations of complex structure on $Y$. In complex dimension $3$, the theory one gets is "Kodaira-Spencer theory of gravity", formulated in \cite{BCOV}. The study of B-model in other dimensions was initiated in \cite{CostelloBCOV}.

\subsection{Large N duality}

A duality, discovered by Gopakumar and Vafa \cite{GV}, relates $G=U(N)$ Chern-Simons theory at level $k$, on 
$$
M= S^3,
$$ 
with A-model topological string on 
$$
X_{{\mathbb P}^1} = {\mathcal O}(-1)\oplus{\mathcal O}(-1) \rightarrow {\mathbb P}^1.
$$
To complete the statement of the duality conjecture, we need to explain the map of parameters, and the correspondence of observables. In defining Chern-Simons theory on the $S^3$, we get to chose two parameters, the integers $N$ and $k$. The Gromov-Witten theory on $X_{{\mathbb P}^1}$ depends on the size of the ${\mathbb P}^1$:
$$
t = \int_{{\mathbb P}^1} \omega
$$
and $\lambda$, the genus counting parameter. The latter enters if, instead of fixing the genus of the Riemann surface $g$, as we did previously, we want to form a generating function, by summing over $g$. The duality maps the parameters of Chern-Simons theory to parameters of Gromov-Witten as follows

$$
t =  {2 \pi  N\over k+N}, \qquad \lambda =  {2 \pi  \over k+N}.
$$
The first prediction of the duality is the equivalence of partition functions {\rm before} we introduce knots in $S^3$:

\begin{equation}\label{fp}
Z_{CS}= Z_{GW}.
\end{equation}
The Chern-Simons partition function on the $S^3$ can be computed as the matrix element of the $S$ matrix (acting on $SU(N)_k$ WZW model on $T^2$)

$$
Z_{CS} (S^3) = \langle 0 | S|0\rangle = S_{00}.
$$
This is an example of obtaining a three manifold, in this case $M'=S^3$, from $M=S^2\times S^1$ by surgery. We start by excising a neighborhood of an unknot in $S^2\times S^1$ running around the $S^1$ and at a point on $S^2$, which splits $M$ into two solid tori. To recover $S^2\times S^1$ we simply glue the the solid tori back together, with trivial identification $U=1$; to obtain an $S^3$ instead, we gluing them with an $S$ transformation of the $T^2$ boundary. 

Gromov-Witten partition function $Z_{GW}$ is defined as the generating function of all maps to $X_{{\mathbb P}^1} $ with no insertions:

$$
Z_{GW} (X_{{\mathbb P}^1} ) = {\rm exp }( \sum_{g=0} \langle 1 \rangle_{g,Q}  \lambda^{2g-2})= {\rm exp }( \sum_{g=0, \beta \in H_2(W)} \langle 1\rangle_{\beta, g} \,Q^{\beta} \lambda^{2g-2} )
$$
where
$$
\langle 1\rangle_{\beta, g} = \int_{[\overline{{\mathcal M}}_{g, 0}(X, \beta)]} 1,
$$
and $Q^\beta = {\rm exp}(- \beta t)$. In this case the degree of the curve is captured by a single number, since $X_{{\mathbb P}^1}$ has a single non-trivial 2-cycle class corresponding to the ${\mathbb P}^1$ itself. The Gromov-Witten partition function of $X_{\mathbb P}^1$ was computed by Faber and Pandharipande in \cite{FP}, by computing $\langle 1 \rangle_{g,Q}$ for every $g$.  The Chern-Simons partition function is known, since the $S$ matrix is known explicitly. Gopakumar and Vafa \cite{GV} showed that $Z_{CS}(S^3)$ equals $Z_{GW}(X_{{\mathbb P}^1})$, by explicit computation. It is striking that the one sums up infinitely many Gromov-Witten invariants in a single matrix element $S_{00}$ in Chern-Simons theory.

The observables of Chern-Simons theory correspond to line operators associated to knots $K$ colored by irreducible representations $R$ of $G$. Introducing a knots on $S^3$ corresponds on $X_{{\mathbb P}^1}$ to allowing maps to have boundaries on a Lagrangian submanifold $L_K$ in $X_{{\mathbb P}^1}$, where $L_K$ gets associated to a knot $K$ in a precise way \cite{OV}. If we have several knots on $S^3$, one will introduce a corresponding Lagrangian for each knot. To explain how these Lagrangians are constructed \cite{Taubes}, we must first explain the origin of  the duality.

\subsubsection{Chern-Simons Theory as a String Theory}
 
$SU(N)$ Chern-Simons theory on a three manifold $M$ turns out to compute \emph{open} topological $A$-model amplitudes on 
$$
X_M = T^*M,
$$
the total space of the cotangent bundle on $M$.
%
One takes the $A$-model topological string on $X_M$, but considers maps with boundaries on $M$:
$$
\phi: \Sigma \rightarrow X_M, \qquad \partial \Sigma \rightarrow M
$$
Allowing boundaries corresponds to considering \emph{open topological A-model}. More precisely, we formally need to take $N$ copies of $M$ in $X_M$,  and keep track of which copy of $M$ a given component of the boundary of $\Sigma$ falls onto. As in the closed case, only the holomorphic maps end up contributing to amplitudes. In fact, as there are no finite holomorphic curves of any kind in $X_M=T^*M$, only degenerate maps contribute -- those where the image curves degenerate to graphs on $M$. Witten showed that the graph expansion that results is the \emph{Feynman graph expansion} of $SU(N)$ Chern-Simons theory. This means that Chern-Simons theory on $M$ computes open topological string amplitudes in this background, in the same way Gromov-Witten theory on $X$ computes closed $A$-model topological string amplitudes on $X$. A mathematical consequence of this is that $G=SU(N)$ Chern-Simons partition function on $M$ must have the following expansion:
%
\begin{equation}\label{pertCS}
Z_{CS}(M) = {\rm exp }( \sum_{g, h=0} F^{CS}_{g,h} N^h {\lambda}^{2g-2+h}),
\end{equation}
where $F^{CS}_{g,h}$ are numbers independent of $N,k$, which capture contributions of maps from surfaces $\Sigma$ that have genus $g$ and $h$ boundary components. For every boundary we have a choice of which copy of $M$ it falls on, hence the power $N^h$, and $\lambda$ keeps track of the Euler characteristic of such as surface which equals $2-2g-h$. 
The numbers $F_{g,h}$ (the perturbative Chern-Simons invariants) play a role in knot theory \cite{vass ,Marino} -- they are related to Vasilliev invariants and to the Kontsevich integral.

Observables in Chern-Simons theory on $M$ are associated to knots. Introducing a knot $K$ in $U(N)$ Chern-Simons theory on $M$ corresponds to, in topological $A$ model on $X_M$, to introducing a Lagrangian $L_K$ which is a total space of the conormal bundle to the knot $K$ in $T^*M$ \cite{OV}. For every point $P$ on the knot $K$ in $M$, one takes the tangent vector to the knot, and defines a rank two sub bundle of the cotangent bundle, by taking all cotangent vectors that vanish on it. The conormal condition implies that $L_K$ is Lagrangian; this in turn guarantees that the adding boundaries preserves topological invariance of the $A$-model. Instead of fixing the representation $R$ coloring the knot, it is better to sum over representations, and consider a formal combination of observables

\begin{equation}\label{OU}
{\mathcal O}_K(U) = \sum_R {\mathcal O}_K(R) \; {\rm Tr}_R U
\end{equation}
where $U$ is an arbitrary unitary matrix of rank $m$, and the sum runs over arbitrary irreducible representations of $U(N)$. The choice of rank $m$ is the number of copies of $L_K$ we take (similarly to the way we took $N$ copies of $M$ to get $SU(N)$ Chern-Simons theory). This observable probes representations $R$ whose Young diagram has no more than $m$ rows, since otherwise ${\rm Tr}_R U$ vanishes. Computing the Chern-Simons partition function in presence of knot $K$ with this observable inserted corresponds to studying $A$-model on $X_{S^3}$ where one allows boundaries on $L_K$. The eigenvalues $(u_1, \ldots, u_m)$ of $U$ keep track of which of the $m$ copies of $L_K$ the boundary component lands on: a single boundary on the $i$-th copy of $L_K$ gets weighted by $u_i$. The resulting partition function is a symmetric polynomial of the $u$'s (using relation between $S_m$ symmetric polynomials and characters of $U(m)$ in various representations) reflecting the $S_m$ permutation symmetry of $m$ copies of $L_K$. The open topological A-model is expected to have the same relation to \emph{open Gromov-Witten} theory, as the closed A-model has to closed Gromov-Witten theory -- where "closed" refers to absence of boundaries of the domain curves $\Sigma$. Unlike the closed Gromov-Witten theory, the foundations of the open Gromov-Witten theory are not entirely in place yet, although progress is being made \cite{Fukaya}.

\subsubsection{Large $N$ Duality is a Geometric Transition}
 
Gopakumar and Vafa conjectured that large $N$ duality has a geometric interpretation, as a transition  
that shrinks the $S^3$ and grows the ${\mathbb P}^1$ :

$$
X_{S^3}\quad  \rightarrow\quad  X_{*} \quad \rightarrow \quad  X_{{\mathbb P}^1}.
$$
In the geometric transition, the $S^3$ disappears and with it the boundaries of maps. If the conjecture is true, it leads to a extraordinary insight: the transition changes topology of the manifolds classically, becomes a change of description -- the theories on $X_{S^3}$  (in presence of boundaries on $N$ copies of the $S^3$) and on 
$X_{{\mathbb P}^1}$ are the same. The passage from one description to the other is perfectly smooth, like a change of charts on a single manifold.

When, $N$ becomes large, it is natural to sum over $h$ in \eqref{pertCS}: while $\lambda$ still keeps track of the Euler characteristic of the underlying Riemann surface, the explicit $N$ dependence disappears. This reflects the fact that both the boundaries and the $S^3$ disapear in the large $N$ dual description. The large $N$ duality implies that
\begin{equation}\label{basicprediction}
\sum_h F^{CS}_{g,h} t^h = F_g^{GW}(t),\qquad t=N\lambda
\end{equation}
which is what Gopakumar and Vafa proved in \cite{GV} by showing $Z_{CS}=Z_{GW}$. We defined $F^{GW}_{g}(t)$ by
$$
Z_{GW}(X_{{\mathbb P}^1}) = {\rm exp }( \sum_{g} F^{GW}_{g}{\lambda}^{2g-2}).
$$ 
\begin{figure}[h]
 \includegraphics[trim = 1mm 90mm 1mm 80mm, clip,
 width=8cm]{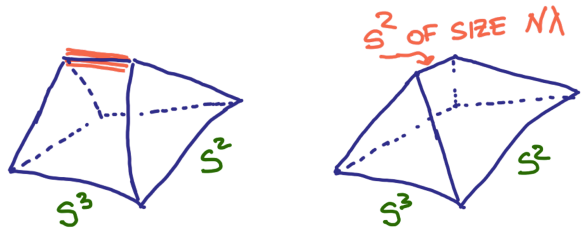}
\caption{Large $N$ duality is a geometric transition that shrinks the $S^3$ and grows an $S^2$ of size $t=N\lambda$.}
\label{four}
\end{figure}

The quantum knot invariants of $K$ are computed by studying open topological $A$ model on $X_{S^3}$ where one allows boundaries on the Lagrangian $L_K$. The geometric interpretation of the large $N$ duality as a transition between $X_{S^3}$ and $X_{{\mathbb P}^1}$ helps us identify what this corresponds on the dual side. 
The asymptotic geometry of $X_{S^3}$ and $X_{{\mathbb P}^1}$ are the same -- they both approach cones over $S^3 \times S^2$ -- they are just filled in differently in the interior. If we first lift $L_K$ off the $S^3$, it can go through the geometric transition smoothly, to become a Lagrangian on $X_{{\mathbb P}^1}$, which we will denote $L_K$ again. In particular, the topology of the Lagrangian is the same,  ${\mathbb R}^2\times S^1$, both before and after the transition. The construction was made precise in \cite{Taubes}. 
\begin{figure}[h]
 \includegraphics[trim = 1mm 80mm 1mm 80mm, clip, width=8cm]{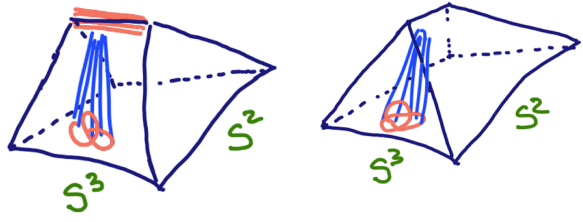}
\caption{The Lagrangian $L_K$ conormal to a knot $K$ in $S^3$ gets pushed through the transition.}
\label{four}
\end{figure}
This leads to a generalization of the basic relation \eqref{basicprediction} as follows: one considers the partition function of Chern-Simons theory, with observable \eqref{OU}  inserted:

\begin{equation}\label{pertCSU}
Z_{CS}(S^3,K, U ) = \langle {\mathcal O}_K(U)\rangle
\end{equation}
This has expansion 

\begin{equation}\label{pertCSU2}
Z_{CS}(S^3,K, U ) = 
 {\rm exp }( \sum_p \sum_{g, h, k_1, \ldots k_p=0} F^{CS}_{g,h, k_1, \ldots k_p } N^h {\lambda}^{2g-2+h+p}\; {\rm Tr} U^{k_1} \ldots  {\rm Tr} U^{k_p}).
\end{equation}
The large $N$ duality conjecture states that one can sum over $h$ to get

\begin{equation}\label{pertCSU2}
Z_{GW}(X_{{\mathbb P}^1}, L_K, U ) = 
 {\rm exp }( \sum_p \sum_{g, k_1, \ldots k_p=0} F^{GW}_{g, k_1, \ldots k_p }(t){\lambda}^{2g-2+p} \; {\rm Tr} U^{k_1} \ldots  {\rm Tr} U^{k_p}),
\end{equation}
in other words, that

\begin{equation}\label{knotprediction}
\sum_h F^{CS}_{g,h, k_1, \ldots k_p } t^h = F^{GW}_{g, k_1, \ldots k_p }(t).
\end{equation}
For simple knots and links -- the unknot and the Hopf-link -- one is able to formulate a computation of the open topological $A$-model amplitude on $X_{{\mathbb P}}^1$ directly in open Gromov-Witten theory \cite{Melissa, MOOP}, and verify the conjecture. For more complicated knots, one needs substantial progress in formulating the open Gromov-Witten side to be able to test the predictions.

The large $N$ duality relating $SU(N)$ Chern-Simons theory to a closed string theory is a part of a family of dualities, whose existence was conjectured by 't Hooft in '70's \cite{Hooft}. He showed that $SU(N)$ gauge theories on general grounds always have Feynman graph expansion of the form \eqref{pertCS}, with coefficients $F_{g,h}$ that depend on the theory, but not on $N$ or $\lambda$. As a consequence, whenever $N$ becomes large, it is natural to re-sum the perturbative series. The result has a form of a closed string Feynman graph expansion; the main question is to identify the dual closed string theory. For Chern-Simons theory on $S^3$, the closed string theory is Gromov-Witten theory on
$X_{{\mathbb P}^1}$. Whenever it exists, the closed string description gets better and better the larger $N$ is, hence the name.

\subsection{Gromov-Witten/Donaldson-Thomas Correspondence}

One is used to studying Gromov-Witten theory by fixing the genus of the Riemann surface $\Sigma$. However, Chern-Simons theory and large $N$ duality suggest it is far more economical to consider all genera at once: Chern-Simons amplitudes are the simplest when written in terms of $q = {\rm exp}(i\lambda)$, rather than $\lambda$. More generally, the fact that one can sum up perturbation series in the gauge theory and solve the theory exactly (at least in principle), is the one of the main reasons why large $N$ duality plays such an important role in physics: the gauge theory allows one to circumvent the usual, genus by genus, formulation of closed string theory.

For Gromov-Witten theory on toric Calabi-Yau three-folds, the theory was indeed solved in this way. 
Using ideas that originated from large $N$ dualities and Chern-Simons theory, \cite{TV} conjectured a solution of Gromov-Witten theory on ${\it any}$ toric Calabi-Yau threefold by cutting up the Calabi-Yau into ${\mathbb C}^3$ pieces, solving the theory on ${\mathbb C}^3$ exactly, and giving a prescription for how to glue the solution on pieces to a solution of the theory on $X$. The result is the topological vertex formalism for Gromov-Witten theory of toric Calabi-Yau manifolds, which expresses the partition function, for a fixed degree $\beta \in H_2(X)$ in terms of rational functions of $q$ \cite{TV}. The topological vertex conjecture was proven by Maulik, Oblomkov, Okounkov and Pandharipande in \cite{MOOP}, who also generalized it away from Calabi-Yau manifolds, to arbitrary toric three-folds. 

The resulting invariants of toric three folds turn out to be directly captured by a precise mathematical theory, Donaldson-Thomas theory of $X$. The theory was introduced in \cite{DT}, and Okounkov, Maulik, Pandharipande and others provided its foundations  \cite{T, DT1, DT2}. The fact that they are also related to Gromov-Witten invariants of $X$, is the content of \emph{Gromov-Witten/Donaldson-Thomas Correspondence} \cite{DT1, DT2}. 
Mathematically, Donaldson-Thomas theory also deals with counting Riemann surfaces in $X$ -- but it does so in a very different way than Gromov-Witten theory. Instead of describing parameterized curves in $X$ in terms of holomorphic maps $\phi:\Sigma \rightarrow X$, as one does in Gromov-Witten theory, in DT theory one describes curves by algebraic equations (see \cite{Okounkov} for a review). Let $X$ be a projective variety, $X\subset {\mathbb P}^r$ for some $r$, and let $z_i$ be the homogenous coordinates of ${\mathbb P}^r$. We can describe the curve $C$ in $X$ as the locus of a set of homogenous polynomials
$$
f(z)=0
$$
which vanish on $C$. The set of all such functions form an ideal $I(C)$ inside ${\mathbb C}[z_0, \ldots z_r]$.
We fix the class $\beta \in H_2(X)$, and $\chi$, the holomorphic Euler characteristic of $C$ ($\chi=1-g$, were $g$ is the arithmetic genus of $C$); and denote the moduli space of $C$ by ${\rm I}(X, \beta, \chi)$. The moduli space is isomorphic to the Hilbert scheme of curves in $X$. $X$ being a threefold is special in this case too: the resulting simplifications allow one to construct a (virtual) fundamental cycle in ${\rm I}(X, \beta, \chi)$, denoted by 
$[{\rm I}(X, \beta, \chi)]$. The analogue of \eqref{qint} is

\begin{equation}\label{qint2}
\langle \gamma_1, \ldots , \gamma_n\rangle_{\beta, \chi} = \int_{[I(X;\beta, \chi)]} \;c_2(\gamma_1)\cdots c_2(\gamma_n)
\end{equation}
To construct $c_2(\gamma)$ takes a special sheaf, the universal ideal sheaf $J$, $J\in {\rm I}(X)\times X$ which has the property that $c_2(J)$ is the locus in ${\rm I}(X)\times X$ corresponding to the set (ideal $I$, point of curve determined by $I$). $c_2(\gamma)$ is the locus of curves meeting $\gamma$ -- this is the coefficient of $\gamma^\vee$ in the decomposition of $c_2(J)\in H^2({\rm I}(X)\times X)$.
Let 
$$Z_{DT}(\gamma, q)_{\beta} = \sum_\chi \langle \gamma_1, \ldots , \gamma_n\rangle_{\beta, \chi}^{DT} \;q^{\chi} 
$$
The conjecture equates this, up to normalization, with 

$$Z_{GW}(\gamma, \lambda)_{\beta} = \sum_\chi  \langle \gamma_1, \ldots , \gamma_n\rangle_{\beta, g}^{GW}\; \lambda^{2g-2}
$$
(In this section we allow disconnected domain curves, as this is the natural thing to do if we want to glue the theory on $X$ from pieces. This is also why we do not exponentiate the right hand sides.) More precisely,

$$
(-\lambda)^{-{\rm vdim}} Z'_{GW}(\gamma; \lambda)_{\beta} = (-q)^{-{\rm vdim}/2}Z'_{DT}(\gamma; q)_{\beta}
$$
where $q={\rm exp}(i\lambda)$, and $'$ denotes dividing by contributions of degree zero curves, which we do on both sides. The Donaldson-Thomas partition function has a beautiful statistical mechanics interpretation in terms of counting boxes stacked up in the toric base of $X$. One sums over a set of box configurations obeying certain natural conditions and weighs the sum with $q^{\# {\rm boxes}}$. Remarkably, the box-counting problem has a saddle point as $q\rightarrow 1$, and $\lambda\rightarrow 0$. In this limit, the cost of adding a box is small and a limiting shape develops, that dominates the partition function $Z_{DT}$ in the limit. Strikingly, the limiting shape encodes the geometry of the Calabi-Yau $Y$ mirror to $X$ \cite{ORV, INOV}. 

The duality relating Gromov-Witten theory and Donaldson-Thomas theory has a physical interpretation in M-theory, a quantum theory that underlies and unifies all string theories \cite{DVV}. Despite the simple appearance -- relating counting curves in two different ways -- the duality that underlies the Gromov-Witten/Donaldson-Thomas correspondence is far from trivial. In particular, Donaldson-Thomas theory leads to many generalizations that go beyond Gromov-Witten theory. In particular, Donaldson-Thomas theory explains the mysterious integrality of Gromov-Witten invariants which was noticed very early on: while one can express Gromov-Witten invariants in terms of a set of integers, this is not manifest from the definition of the theory -- Gromov-Witten theory naturally leads to counts of curves which are rational numbers, not integers, since the underlying moduli spaces are not smooth. One expects that relation of Donaldson-Thomas and Gromov-Witten theories is much like the diagram in Fig.7 -- there is a large parameter space of DT theory, the tips of which have Gromov-Witten interpretation. 
\section{Combining dualities and knot theory}

Duality is like a change of charts on a manifold; in particular, we can combine dualities, and get even more mileage from them. For example, one can \emph{combine large $N$ duality and mirror symmetry}. It turns out that this can shed fundamentally new light on knot theory, but to explain this we need to back up to explain the origin of mirror symmetry first. 

\subsection{Homological Mirror Symmetry and the SYZ Conjectures}
We have seen that Gromov-Witten theory computes quantum corrections to the classical geometry of a Calabi-Yau $X$. Mirror symmetry sums up these corrections, in terms of the geometry of the mirror Calabi-Yau $Y$. One can make this precise, and give a (conjectural) description for how the classical geometry of $Y$ emerges from quantum geometry of $X$.

There are two mathematical conjectures that capture aspects of mirror symmetry. \emph{Homological mirror symmetry conjecture} of Kontsevich \cite{K} relates categories of allowed boundary conditions of topological A-model on $X$ and topological $B$ model on $Y$. The former is captured by the Fukaya category ${D^{\mathcal F}(X)}$ of $(X, \omega)$ whose objects are Lagrangian submanifolds $L\subset X$ equipped with a unitary flat connection $A$:
$$
\omega|_L = 0, \qquad F=0,
$$
where $F=dA$ is the curvature of the flat connection $A$. Holomorphic maps $\phi: \sigma \rightarrow X$ are allowed to have boundaries on $L\in X$; the connection on $L$ couples to the boundaries. We have seen examples of this, when $X=T^*M$, where we took $L=M$ or $L_K$, the Lagrangian associated to the knot. The $A$ connection on $M$ is the Chern-Simons connection. The morphisms in the category are associated to strings with endpoints on pairs of Lagrangians. Kontsevich conjectured that on the mirror $Y$ there is an equivalent category, the bounded derived category of coherent sheaves, $D^b(Y)$.  Homological mirror symmetry conjecture was recently proven for a famous example of the quintic Calabi-Yau manifold $X$ and its mirror \cite{sheridan}.

Among the objects in $D^b(Y)$, a privileged role is played by the structure sheaf ${\mathcal O}_p$, for $p$ a point in $Y$. The moduli space of ${\mathcal O}_p$ is $Y$ itself. Mirror symmetry implies that there must be an object in the Fukaya category of $X$ with the same moduli space.  
Strominger, Yau and Zaslow \cite{SYZ} showed that this fact alone implies that the mirror pair of manifolds $(X,Y)$ must both be $T^3$ fibrations over a common base $B$, with fibers that are (special) Lagrangian tori. Let $X$ be a $T^3$ fibration, 
$$
T^3 \rightarrow X \rightarrow B
$$
over a base $B$,  and $L_p$ be a $T^3$ fiber of $X$ above a point in $p\in B$. The moduli space of $L_p$  is the base $B$ itself. The full moduli space is a fibration over this, by moduli of a flat $U(1)$ bundle on $T^3$. The moduli of a $U(1)$ bundle on $T^3$ is the dual torus ${\hat T}^3$. More precisely, the resulting moduli can get corrected by "disk instantons" -- maps from the disk to $X$ with boundaries on $L$, and taking this into account results in the mirror manifold:
$$
{\hat T}^3 \rightarrow Y \rightarrow B.
$$
This is the \emph{SYZ mirror symmetry conjecture}. This gives a simple geometric picture of mirror symmetry, explicitly constructing the mirror $Y$ from the quantum moduli space of objects on $X$.   The duality that relates string theory on a circle $S^1$ of radius $R$ to a string theory on a dual circle ${\hat S}^1$ of radius $1/R$  (or a product of circles), is a very basic example of a duality in string theory, called $T$-duality. Here, we see that mirror symmetry is simply $T$-duality, applied fiber-wise, over each point in $B$. For a review of SYZ conjecture, see \cite{SYZr}.

\begin{figure}[h]
 \includegraphics[trim = 1mm 90mm 1mm 90mm, clip,
 width=10cm]{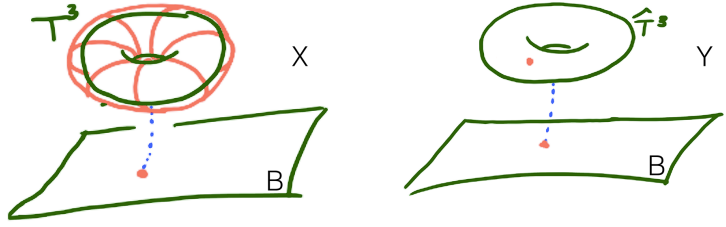}
\caption{SYZ Mirror Symmetry}
\label{five}
\end{figure}

One can extend the SYZ conjecture away from compact Calabi-Yau manifolds. When $X$ is a toric Calabi-Yau manifold, it is non-compact, and then the $T^3$ fibration is replaced by an $T^2\times {\mathbb R}$ fibration over $B= {\mathbb R}^3$. The geometry of the toric Calabi-Yau is, from this perspective, encoded in the geometry of a trivalent graph $\Gamma$ in $B$ over which the $T^2$ fiber degenerates to $S^1$. The role of $L_p=T^3$ in the compact case is replaced by $L_p= S^1\times {\mathbb R^2}$: the generic $T^2 \times {\mathbb R}$ fiber degenerates, over $\Gamma$ to a union of two copies of $L_p$, and we take one of them.  The classical moduli space of $L_p$ is the graph $\Gamma$; the moduli of the flat connection on $L_p$ is a circle fibered over this, and one still has to take into account disk instanton corrections. The quantum moduli space is a Riemann surface
\begin{equation}\label{mirrR}
H_{\Gamma}(x,p) =0 , \qquad x,p \in {\mathbb C}^{*}
\end{equation}
The mirror Calabi-Yau is a hypersurface
\begin{equation}\label{mirrC}
Y:\qquad uv - H_{\Gamma}(x,p) =0 , \qquad u,v\in {\mathbb C}
\end{equation}
Mirror to $L_p$ is no longer a structure sheaf on $Y$, but instead a sheaf supported on a curve, corresponding to choosing a point on the mirror Riemannn surface \eqref{mirrR} and picking either $u=0$ or $v=0$, depending on which component of the reducible $T^2\times {\mathbb R}$ fiber we took.

Now, let us describe what this has to do with knot theory.

\subsection{Large $N$ Duality, SYZ Mirror Symmetry and Knot Theory}

Large $N$ duality relates $SU(N)$ Chern-Simons theory on $S^3$ to Gromov-Witten theory (or A-model topological string) on
$X_{{\mathbb P}^1}$; this is a non-compact, toric Calabi-Yau manifold. We can obtain its mirror by application of SYZ mirror symmetry, by finding the quantum moduli space of a Lagrangian in $X_{{\mathbb P}^1}$ of topology of ${\mathbb R}^2\times S^1$. 

In fact, we get one such Lagrangian for \emph{every knot}  $K$ in $S^3$, and with it a distinct mirror $Y_K$ \cite{AV1}. To construct a Lagrangian $L_K$ in $X_{{\mathbb P}^1}$ of topology of ${\mathbb R}^2\times S^1$ corresponding to a knot $K$ in $S^3$, one starts with a Lagrangian in $X_{S^3} = T^*S^3$, which is a total space of the conormal bundle to the knot $K$ in the $S^3$ base, lifts it off the zero section (so that it does not intersect the singular locus when the $S^3$ shrinks), and then pushes it through the transition that relates $X_{S^3} $ and $X_{{\mathbb P}^1}$. The mirror depends only on the homotopy type of the knot:

\begin{equation}\label{mirrCK}
Y_K:\qquad uv - H_{K}(x,p) =0.
\end{equation}
The quantum moduli space of $L_K$ in $X_{{\mathbb P}^1}$ is the Riemann surface
\begin{equation}\label{mirrHK}
H_K(x,p)=0.
\end{equation}
The pair $x$ and $p=p_K(x)$ that lie on \eqref{mirrHK} are determined by summing holomorphic disks with boundaries on $L_K$.
Large $N$ duality \eqref{knotprediction} in turn relates this to a limit of corresponding Chern-Simons amplitude:
$${\rm log}\, p_K(x) = x{d\over dx} {\rm lim}_{\lambda \rightarrow 0}\; \lambda \,  \langle {\mathcal O}_K(x)\rangle
$$
 where one takes takes $U=x$ to be a rank one matrix, and $Q={\rm exp}(-t)$. For example, taking the knot $K$ to be the unknot, one gets the "conventional" mirror of $X_{{\mathbb P}^1}$, where 
$$H_{\bigcirc}(x,p) =1- x- p + Q xp.$$
But, taking $K$ to be a trefoil knot instead, as an example, we get a different answer:
$$H_{K}(x,p) =1- Qp + (p^3-p^4+2 p^5- Qp^6+Q^2 p^7) x - (p^9-p^{10})x^2.$$

Thus, the combination of two string dualities, large $N$ duality and mirror symmetry, gives rise to a new knot invariant, the mirror Calabi-Yau manifold $Y_K$. Chern-Simons theory produces an infinite list of knot invariants, differing by the representations coloring the knot. To tell knots apart, it is necessary, though maybe not sufficient, to compare the entries of this list. String duality suggests that one can replace the entire list with a single invariant, the mirror Calabi-Yau manifold $Y_K$, plus presumably a finite set of data needed to define the quantization in this setting. Once the quantization procedure is defined, topological B-model string is a functor, that associates to $Y_K$ quantum invariants. Moreover, unlike knots, Calabi-Yau manifolds are easy to tell apart, simply by comparing the polynomials $H_K(x,p)$. Thus, instead of quantum physics playing the central role in constructing good knot invariants, classical geometry of $Y_K$ becomes the key.

The Riemann surface $H_K(x,p)=0$ turns out to have an alternative mathematical formulation, as the augmentation variety of the knot \cite{AV2}. This is one of the knot invariants that arise from knot contact homology. Knot contact homology is the open version, developed by Lenhard Ng \cite{Ng}, of the symplectic field theory approach to counting holomorphic curves, pioneered by Eliashberg and Givental. This provides a relation between two distinct approaches to counting holomorphic curves, one coming from Gromov-Witten theory, and the other from symplectic field theory.

\section{M-theory and Homological knot invariants}

There is a mysterious aspect of Chern-Simons knot invariants. From the definition of the Jones polynomial $J_K(q)$, one can see that it is always a Laurent polynomial in $q^{1/2} $ with {\it integer} coefficients.  Coefficients of the knot polynomials are always integers, as if they are counting something. What are they counting? Since $q=e^{i\lambda}$, where $\lambda$ is either the Chern-Simons or the topological string coupling constant, the answer to this question cannot come from Chern-Simons theory or topological string.

Khovanov made this structure manifest in a remarkable way. He constructed a bi-graded homology theory, in such a way that the Jones polynomial arises as the Euler characteristic

$$J_K(q) = \sum_{i,j} \,(-1)^j \,q^{i/2} \,{\rm dim }\,{\mathcal H}_{i,j}(K),
$$
counting dimensions of knot homology groups,
$$
{\mathcal H}_{i,j}(K),
$$
with signs. The Poincare polynomial of knot homology
$$
P_K(q,t) = \sum_{i,j} t^j q^{i/2} {\rm dim }\,{\mathcal H}_{i,j}(K)
$$
has strictly more information about the knot, it is a better knot invariant. One expects that this should have generalizations to all Chern-Simons (Witten-Reshetikhin-Turaev) knot and three manifold invariants, however knot homology theories are extremely complicated. A unified approach to categorification of quantum group invariants was very recently put forward in \cite{Webster1, Webster2}. As far as we are aware, a fully combinatorial construction of knot homologies is available only for the Jones polynomial itself. 

Knot homologies have a physical interpretation within M-theory \cite{ GSV, W}, due to Gukov, Vafa and Schwarz, and later Witten. Knot homologies are Hilbert spaces of states which preserve some supersymmetry in M-theory realization of Chern-Simons theory. To obtain Chern-Simons theory from M-theory, one uses a similar geometry as in topological string theory. Witten was able to reduce the M-theory construction to computing cohomologies of spaces of solutions to a certain equation, the Kapustin-Witten equation \cite{W, W2}, with boundary conditions depending on the knot type, but mathematics and physics are still comparably complex. It was shown in \cite{WittenGaiotto} that the approach of \cite{W} leads to the Jones polynomial, once one computes the Euler characteristic. However, this is yet to lead to an explicit construction of knot homologies and $P_K(q,t)$, even in examples. 

Physics does provide a powerful insight, if one restricts to three-manifolds and knots respecting a certain circle symmetry. In the presence of the extra symmetry, one can formulate, using M-theory, a three dimensional topological theory, the \emph{refined Chern-Simons theory} \cite{AS}. The partition function of the refined Chern-Simons theory (conjecturally) computes a two-variable polynomial $I_K(q,t)$, a close cousin of Poincare polynomial of homology theory categorifying Chern-Simons theory. For three manifolds and knots admitting a (semi-free) circle action, knot homologies corresponding to arbitrary ADE gauge groups and their representations, should admit an additional grade:
$$
{\mathcal H}_{i,j}(K) = \oplus_k {\mathcal H}_{i,j,k}(K).
$$ 
This leads to an index, more refined than the Euler characteristic:

$$
I_K(q,t) = \sum_{i,j, k} (-1)^k q^{i/2} t^{j+k} {\rm dim }\,{\mathcal H}_{i,j, k}(K),
$$
akin to the Hirzebruch $\chi_y$ genus. Setting $t=-1$, both $P_K(q,t)$ and $I_K(q,t)$ reduce to Chern-Simons invariants.

The refined Chern-Simons theory, which computes $I_K$, is solvable explicitly. As in Witten's solution of the "ordinary" Chern-Simons theory -- by cutting the three-manifold into pieces, solving the theory on pieces and gluing -- one reduces the problem of computing the knot and three manifold invariants to matrix multiplication. In fact, since refined Chern-Simons theory exists for a restricted set of three-manifolds and knots (those admitting a circle symmetry), a smaller set of ingredients enter -- all one needs are the $S$ and the $T$ matrix providing a representation of $SL(2,Z)$ on the Hilbert space ${\mathcal H}_{T^2}$. The $S$ and $T$ matrices now depend on both $q$ and $t$ (they are given in Macdonald polynomials of the corresponding ADE group, evaluated at a special point, generalizing the Schur polynomials in Chern-Simons case.) This is immeasurably simpler than constructions of homologies themselves. Even better, for simple representations of $SU(N)$, at large $N$ (corresponding to categorification of the HOMFLY polynomial), the index $I_K$ and the Poincare polynomial $P_K$ of knot homology theory agree. This gives strong evidence that refined Chern-Simons theory indeed computes a new genus on knot homologies, and also evidence that M-theory is indeed behind knot homologies.

It is striking that, even though the refined Chern-Simons theory has been formulated only recently, many connections have already been made. It is known that refined Chern-Simons invariants are related to $q$-deformation of conformal blocks of $W$-algebras \cite{AS2}; they have deep connections to the K-theory of the Hilbert scheme of points on ${\mathbb C}^2$ \cite{Shende, Nakajima, Negut}. The knot invariants arising from refined Chern-Simons theory have a direct connection to representation theory of Double Affine Hecke Algebras (DAHA) \cite{Negut}. There is evidence that the invariants are also related to Donaldson-Thomas invariants of toric three-folds constructed recently in \cite{NO}. 

\section{Outlook}
Despite the successes of string theory in solving difficult problems in mathematics, this is no doubt just a tip of the iceberg. All string theories are unified in a single theory, M-theory. Genus by genus expansion, on which topological string and superstring theories are based, exists only at the corners of M-theory parameter space. Dualities fill in the rest of the diagram. M-theory has already made an appearance in knot theory context, and in relating Gromov-Witten theory to Donaldson-Thomas theory.

\begin{figure}[h]
 \includegraphics[trim = 1mm 20mm 1mm 20mm, clip,
 width=7cm]{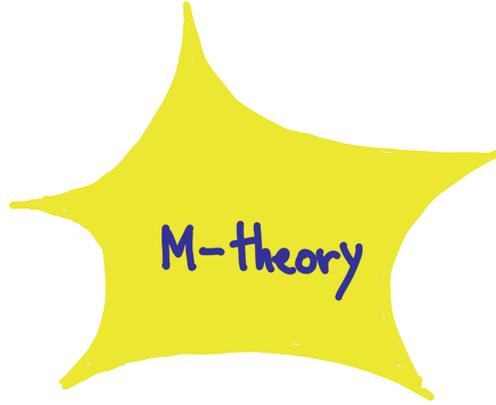}
\caption{M-theory is believed to be the unique quantum theory underlying all string theories. Different descriptions of it, which emerge at the corners of the diagram, are related by dualities.}
\label{six}
\end{figure}
Mathematical consequences of dualities in M-theory are largely unexplored. Two topological string theories, the A- and the B-model, with their many mathematical uses, capture supersymmetric M-theory partition functions in a very specific background \cite{GVM1,GVM2}. The plethora of mathematical predictions extracted from topological string and its dualities, such as mirror symmetry and large $N$ duality we described provide just a glimpse of the mathematical content of M-theory. Supersymmetric partition functions of M-theory are generalizations of topological string, yet only their exploration has only just begun, see \cite{NO}.

To be sure, dualities do not require string theory. There are examples of dualities in quantum gauge theories which can be stated without invoking string theory. Even so, string theory often plays the crucial role in discovering the dualities, and in studying them. \emph{Symplectic duality}  \cite{symplectic}, which plays an important role in knot theory and other areas of mathematics, is a duality of quantum gauge theories in three dimensions. Even though today one can phrase it purely in gauge theory language, the duality was discovered using string theory,  in \cite{HananyWitten, IS}, and string theory helps one understand the it better \cite{GW}. \emph{Seiberg-Witten (SW) theory}, the celebrated 4d QFT with an important role for 4-manifold invariants \cite{SW, WS}, turns out to have many dual descriptions \cite{Gaiotto:2009we}. In fact most of the theories in this class turn out not have a conventional description, but need M-theory for their definition. To define them, one considers the 6-dimensional the "theory $X$" that arises as a part of M-theory, compactified on a Riemann surface $C$.  Only in certain corners of the moduli of $C$ the usual gauge theory description emerges. This observation leads to a precise mathematical prediction: the partition functions of this class of Seiberg-Witten theories are the conformal blocks on $C$ of a class of 2d conformal field theories with $W$-algebra symmetry \cite{AGT}. This unifies problems in QFT, geometry and representation theory. Some aspects of this correspondence were recently proven by \cite{Nakajima2}. The \emph{S-duality} of 4d ${\mathcal N}=4$ Yang-Mills theory, related to electric-magnetic duality, is believed to be the duality underlying the geometric Langlands program \cite{MonO,VW, Kapustin:2006pk, Donagi}. The Langlands program has, for the last 50 years, been one of the key unifying themes in mathematics \cite{Frenkel}. Once again, even though one can phrase $S$ duality in terms of gauge theory alone, much of our understanding of it comes from string theory: the 4d ${\mathcal N}=4$ Yang-Mills theory arises by compactifying the 6d theory $X$ on a torus, and S-duality simply comes from $SL(2,Z)$ symmetry of the torus!

The interacting between the two fields has only really begun in ernest. It is fairly certain that dualities in string theory and quantum field theory hold potential for many new breakthroughs in mathematics, by extracting their mathematical predictions, and proving them. It should also lead to a deeper and sharper understanding of quantum physics. There is a good chance that eventually, our view of mathematics, and quantum physics will have changed profoundly.

\section{Acknowledgments}

I am grateful to Robbert Dijkgraaf, Tobias Ekholm, Nathan Haouzi, Albrecht Klemm, Markos Marino, Lenhard Ng, Shamil Shakirov, Andrei Okounkov, Cumrun Vafa for years of collaborations that taught me the ideas presented, and the numerous mathematicians and physicists who have worked together to unearth the beautiful structures in our subject.


\end{document}